\documentclass{ws-rv9x6}
\usepackage{subfigure}   
\usepackage{ws-rv-thm}   
\usepackage{ws-rv-van}   
\usepackage{amsmath,mathrsfs,dsfont}
\usepackage{amsfonts}
\makeindex
\newcommand{\be}{\begin{equation}}
\newcommand{\ee}{\end{equation}}
\newcommand{\beqa}{\begin{eqnarray}}
\newcommand{\eeqa}{\end{eqnarray}}

\begin{document}

\centerline{\Large{Noncommutative $AdS_2$ I: Exact solutions}}

\vspace{5mm}

\author[A. Pinzul]{Aleksandr Pinzul}

\address{Universidade  de Bras\'{\i}lia, Instituto de F\'{\i}sica\\
and\\
International Center of Physics\\
70910-900, Bras\'{\i}lia, DF, Brasil\\
apinzul@unb.br}

\author[A. Pinzul and A. Stern]{Allen Stern}

\address{Department of Physics, University of Alabama,\\ Tuscaloosa,
Alabama 35487, USA, \\
astern@ua.edu}
{\it In Honor of A.P.Balachandran on the Occasion of His 85th Birthday}
\vspace{5mm}

\begin{abstract}
We study the exact solutions of both, massless and massive, scalar field theory on the noncommutative $AdS_2$. We also discuss some important limits in order to compare with known results.
\end{abstract}


\body


\section{Introduction}\label{ra_sec1}
The study of noncommutative spaces and field theories on them has a great number of motivations and applications, ranging from Quantum Hall Effect \cite{Susskind:2001fb} to Superstring Theory \cite{Seiberg:1999vs}. This, of course, is in addition to the purely mathematical interest in noncommutative geometry \cite{Connes:1994yd}. Of special interest is the study of noncommutative spaces that preserve the isometries of their commutative counterparts, one of the most famous examples being the so-called fuzzy sphere \cite{Madore:1991bw}. Such spaces serve as good models to simulate the ultra-violet properties of field theories due to universal quantum gravity effects (see \cite{Doplicher:1994zv,Doplicher:1994tu} for a general model independent argument for appearance of the space-time noncommutativity as a quasiclassical quantum gravitational effect and, e.g., \cite{Subjakova:2020haa} for a review of some recent studies of a scalar field theory on a fuzzy sphere).

In this paper we study the exact solutions to the dynamics of a free scalar field on noncommutative (Euclidean) $AdS_2$ ($ncAdS$). This space is the non-compact cousin of the fuzzy sphere. One of the main motivations to study such a model, aside from those briefly mentioned above, is its importance to the celebrated $AdS/CFT$ correspondence \cite{Maldacena:1997re}, specifically to the 2-dimensional version, $AdS_2/CFT_1$. If one believes that the correspondence holds in the strong sense, i.e. including at the quantum gravity scale, then, from the observation that noncommutativity models quasiclassical quantum gravitational effects, it follows that there should be some form of the correspondence between a field theory on $ncAdS$ and some conformal theory on the one-dimensional boundary of the $AdS$ space. This strategy was adopted in \cite{Pinzul:2017wch,deAlmeida:2019awj} where strong perturbative evidence in favor of such a correspondence was found. The results of \cite{Lizzi:2020cwx} suggest that this also should be true in higher dimensions. The work \cite{Pinzul:2021cjz} was the first attempt to go beyond the perturbative (in the noncommutative parameter) analysis. For the case of a free massless scalar field, the exact solutions for the dynamics were found and it was shown that the exact 2-point correlator of boundary fields obtained via the $AdS/CFT$ correspondence prescription is conformally invariant.

In this paper, we develop another approach to construction of the exact solutions of a free scalar theory on $ncAdS$, which allows for an extension of the results \cite{Pinzul:2021cjz} to the massive case. This, in turn, allows for a more general test of the correspondence. This development is reported in another contribution to the volume \cite{Pinzul:2022}.

The plan of the paper as follows. In section \ref{The model} we introduce our model, as well as give some technical details. The general construction of the exact solutions is done in section \ref{Exact solutions}, while section \ref{Special cases} discusses some important limits, as well as reproduces the results of \cite{Pinzul:2021cjz} for the massless case. In section \ref{Discussion} we discuss possible future developments of the project.

\section{The model}\label{The model}
In this section, we briefly introduce our model. For more details, see \cite{Pinzul:2017wch,deAlmeida:2019awj,Pinzul:2021cjz}.

Noncommutative Euclidean $AdS_2$ is obtained by quantizing the $SU(1,1)$ invariant Poisson structure of commutative $AdS_2$. Commutative Euclidean $AdS_2$ ($EAdS_2$) can be defined in terms of imbedding coordinates $X^a$, $a=0,1,2$, spanning $\mathbb{R}^{1,2}$ and  satisfying
\be\label{constrain}
\eta_{ab}X^a X^b = -1 \ ,
\ee
where the metric on $\mathbb{R}^{1,2}$ is $[\eta_{ab}] =$ diag$(-1, +1, +1)$. The corresponding $SU(1,1)$-invariant Poisson structure is given by
\be\label{Poisson1}
\{X^a , X^b\} = \epsilon^{abc} X_c \ ,
\ee
where rasing/lowering of indices is done with the help of $\eta_{ab}$ and $\epsilon^{abc}$ is totally antisymmetric with $\epsilon^{012} =1$. The field equations for a free massive scalar field $\Phi$ on $EAdS$ can be compactly written in terms of this structure:
\be\label{eomcomm}
\{X^a ,\{X_a , \Phi\}\} = m^2 \Phi \ .
\ee

To study the $AdS/CFT$ correspondence, it is convenient to also introduce local coordinates. The so-called Feffermann-Graham coordinates, $(t,z)$, can be defined in terms of the imbedding coordinates as\footnote{Here we restrict to the single sheeted hyperboloid with $X^0 \le -1$, which implies $z>0$ and so $(t,z)$ span the half-plane.}
\be\label{FG}
z = \frac{1}{X^2 - X^0}\ ,\ \ t = -\frac{X^1}{X^2 - X^0}\equiv -z X^1
\ee
and the Poisson structure (\ref{Poisson1}) takes the form
\be\label{Poisson2}
\{t,z\} = z^2 \nonumber\ .
\ee

The quantization of the model, preserving the constraint (\ref{constrain}), is rather straightforward (see \cite{Pinzul:2017wch,deAlmeida:2019awj,Pinzul:2021cjz} for details). Essentially, it amounts to promoting the imbedding coordinates $X^a$ to operators $\hat{X}^a$ satisfying the quantized version of the Poisson structure
\be
[\hat{X}^a , \hat{X}^b] = i\alpha\epsilon^{abc} \hat{X}_c \nonumber\ ,
\ee
where $\alpha$ is the parameter of noncommutativity, which is related to the quantum gravitational scale. The noncommutative version of the scalar field equation (\ref{eomcomm}) becomes \cite{Pinzul:2021cjz}
\be\label{XfyXeqmfy}
\frac{\alpha^2}2\hat \Delta\hat \Phi=
\hat X_{a}\hat \Phi\hat X^{a}\,+\,\hat\Phi\;=\; \frac{\alpha^2m^2}2\hat \Phi \ ,
\ee
where $\hat \Delta := [\hat{X}^a , [\hat{X}_a, \cdot\ ]]$ is the noncommutative Laplacian.

\section{Exact solutions}\label{Exact solutions}
To find the solutions of (\ref{XfyXeqmfy}), we will first re-write (\ref{XfyXeqmfy}) in terms of the quantized Feffermann-Graham coordinates, $(\hat t, \hat z)$, which are given by the quantization of (\ref{FG})\footnote{In \cite{Pinzul:2017wch} we showed that $\hat{X}^2 - \hat{X}^0$ has a positive spectrum for the discrete series irreducible representation of $su(1,1)$, which exactly corresponds to the noncommutative Euclidean $AdS$.}
\be\label{FGquant}
\hat{z} = \frac{1}{\hat{X}^2 - \hat{X}^0}\ ,\ \ \hat{t} = -\frac{1}{2}(\hat{z} \hat{X}^1 + \hat{X}^1 \hat{z})\nonumber\ .
\ee
The only nontrivial step is the inversion of these relations to obtain $\hat{X}^a$ as a function of $\hat z$ and $\hat t$. It is more convenient to formally introduce the quantized ``radial'' coordinate, $\hat r = \hat{z}^{-1}$, satisfying a simple, canonical, commutation relation
\be\label{hatthatzinv}
[\hat{t},\hat{r}]= - i\alpha \ .
\ee
Then the noncommutative imbedding coordinates can be written in terms of the local ones as \cite{Pinzul:2021cjz}
\beqa\label{Xtz}
&&\hat{X}^0 = -\frac{1}{2}(\kappa^2 \hat{r}^{-1} + \hat{t}\hat{r}\hat{t} + \hat{r})\ ,
\hat{X}^1 = -\frac{1}{2}(\hat{r}\hat{t} + \hat{t}\hat{r}) \ ,\nonumber \\
&&\hat{X}^2 = -\frac{1}{2}(\kappa^2 \hat{r}^{-1} + \hat{t}\hat{r}\hat{t} - \hat{r})\ ,
\eeqa
where the nontrivial deformation parameter, $\kappa=\sqrt{1+\frac{\alpha^2}4}$, is necessary to preserve the quantum version of (\ref{constrain}).

Using (\ref{Xtz}) in (\ref{XfyXeqmfy}), one gets
\beqa
2{\alpha^2} \hat \Delta\hat \Phi&=&-(\kappa^2 \hat{r}^{-1} + \hat{t}\hat{r}\hat{t} + \hat{r})\hat\Phi(\kappa^2 \hat{r}^{-1} + \hat{t}\hat{r}\hat{t} + \hat{r})\cr
&&+(\hat{r}\hat{t} + \hat{t}\hat{r})\hat \Phi(\hat{r}\hat{t} + \hat{t}\hat{r}) \cr
&& + (\kappa^2 \hat{r}^{-1} + \hat{t}\hat{r}\hat{t} - \hat{r})\hat\Phi(\kappa^2 \hat{r}^{-1} + \hat{t}\hat{r}\hat{t} - \hat{r}) + 4\hat\Phi\cr
&=&(\hat{r}\hat{t} + \hat{t}\hat{r})\hat \Phi(\hat{r}\hat{t} + \hat{t}\hat{r})-2\left(\hat t\hat{r}\hat t\hat \Phi\hat{r}+\hat{r}\hat \Phi\hat t\hat{r}\hat t\right)\cr
&& -2\kappa^2\,\left(\hat{r}^{-1}[\hat \Phi,\hat{r}]+[\hat{r},\hat \Phi]\hat{r}^{-1}\right) -\alpha^2\hat\Phi \nonumber\ .
\eeqa
This can be simplified further by using (\ref{hatthatzinv})
\beqa
{\alpha^2}\hat \Delta\hat\Phi
&=& - [\hat{t},\hat{r}[\hat{t},\hat\Phi]\hat{r}] - \kappa^2 \hat{r}^{-1}[\hat{r},[\hat{r},\hat\Phi]]\hat{r}^{-1} \nonumber\ ,
\eeqa
so the full noncommutative dynamics will be given by
\be
- [\hat{t},\hat{r}[\hat{t},\hat\Phi]\hat{r}] - \kappa^2 \hat{r}^{-1}[\hat{r},[\hat{r},\hat\Phi]]\hat{r}^{-1} = \alpha^2 m^2 \hat\Phi \nonumber
\ee
or
\be\label{eom}
- \hat{r}[\hat{t},\hat{r}[\hat{t},\hat\Phi]\hat{r}]\hat{r} - \kappa^2 [\hat{r},[\hat{r},\hat\Phi]] = \alpha^2 m^2 \hat{r}\hat\Phi\hat{r} \ .
\ee

To further simplify the equation of motion, we note that a trivial consequence of the commutation relation (\ref{hatthatzinv}) is
\beqa\label{derivatives}
&&[\hat{r},T ] = i\alpha \dot{T}\ , \ [\hat{t},R ] = -i\alpha R'\ ,
\eeqa
where $T=T(\hat{t})$, $R=R(\hat{r})$, $\dot{T} = \frac{d}{dt}T(t)|_{t=\hat{t}}$ and $R' = \frac{d}{dr}R(r)|_{r=\hat{r}}$.
Then it is easy to see that the following ansatz for $\hat\Phi$
\be\label{ansatz}
\hat \Phi=T({\hat{t}})R(\hat{r})T({\hat{t}}) \ \ \mathrm{with}\ \ \dot{T} = i\beta T \ ,
\ee
for any $\beta$ (it will be necessary to restrict to real $\beta$, see below), separates variables in (\ref{eom}) in the following sense. Plugging (\ref{ansatz}) into (\ref{eom}) and using (\ref{derivatives}), one immediately gets
\beqa\label{separ1}
T (\hat{r} - \alpha\beta)\frac{d}{dr}\Big((\hat{r} - \alpha\beta)R'(\hat{r}+ \alpha\beta)\Big)(\hat{r} + \alpha\beta)T - 4\kappa^2 \beta^2 TRT\nonumber\\
= m^2 T(\hat{r} - \alpha\beta)R(\hat{r} + \alpha\beta)T \ .
\eeqa
Using that $T$ is invertible (actually from (\ref{ansatz}), $T = \exp(i\beta\hat{t})$) and any two functions of $\hat{r}$ commute, (\ref{separ1}) can be written as\footnote{We also have to assume that, at least formally, $\alpha^2\beta^2 - \hat{r}^2$ is invertible too. This could be achieved by adding an infinitesimal imaginary part, which will amount to choosing the correct analytic continuation, see below.}
\be\label{eom1}
\left[(\alpha^2\beta^2 - \hat{r}^2)R'\right]' + \left[m^2 - \frac{4\kappa^2\beta^2}{\alpha^2\beta^2 - \hat{r}^2}\right]R = 0 \ .
\ee

For $\alpha\beta\ne 0$ this is essentially the general Legendre equation:
\be
\left[(1 - x^2)L'\right]' + \left[\nu(\nu +1) - \frac{\mu^2}{1 - x^2}\right]L = 0 \nonumber
\ee
that has eight standard solutions (of course, only two of them being linearly independent). For the case of (\ref{eom1}) the solutions can be written as \cite{Bateman:100233}
\beqa\label{solutions}
&&R_{1\pm} = P_{\nu-\frac 12 }^{\pm\frac 2\alpha\kappa}\left(\frac{\hat{r}}{\alpha  \beta}\right)\;,\quad R_{2\pm} = Q_{\nu-\frac 12 }^{\pm\frac 2\alpha\kappa}\left(\frac{\hat{r}}{\alpha  \beta}\right)\;, \nonumber\\
&&R_{3\pm} = P_{-\nu-\frac 12 }^{\pm\frac 2\alpha\kappa}\left(\frac{\hat{r}}{\alpha  \beta}\right)\;,\quad R_{4\pm} = Q_{-\nu-\frac 12 }^{\pm\frac 2\alpha\kappa}\left(\frac{\hat{r}}{\alpha  \beta}\right)\ ,
\eeqa
where $\nu=\sqrt{ m^2+\frac 14}$.

For $\alpha=0$, after change of the variable (which now can be thought as the commutative one), $r = 1/z$, and substituting $R = \sqrt{z}\tilde{R}$, equation (\ref{eom1}) becomes the modified Bessel's equation
\be
z^2 \frac{d^2}{dz^2}\tilde{R} + z \frac{d}{dz}\tilde{R} - (m^2 + \frac{1}{4} + 4\beta^2 z^2)\tilde{R} = 0 \nonumber\ ,
\ee
which leads to the well-known commutative solution \cite{Freedman:1998tz}
\be
R = \sqrt{z}  K_\nu\left({2 \beta z }\right)\;,\qquad\nu=\sqrt{ m^2+\frac 14}\ ,\label{cmtvsln}
\ee
where $K_\nu$ is the modified Bessel function of the second type (the other solution given in terms of the modified Bessel function of the first type, $I_\nu$, exponentially blows up when $z\rightarrow\infty$, so it does not satisfy the asymptotic regularity condition).

In order to better understand the above results, we next examine some important limits.

\section{Special cases}\label{Special cases}
Here we would like to study two important limits of the general result (\ref{solutions}): the commutative limit and the massless one.

\subsection{Commutative limit}
A more careful study of the commutative limit is necessary to understand which of the solutions in (\ref{solutions}) should be taken in the fully noncommutative case. Formally, the limit corresponds to sending $\alpha$ to zero. This limit is somewhat tricky because it corresponds to simultaneously sending both, the upper index and the argument, to infinity in either of the solutions (\ref{solutions}). It is more convenient to do this using the integral representations. By doing so, we will show that the correct choice, i.e. the one that has the correct commutative limit (\ref{cmtvsln}), is $R_{1-}$.

The well-known integral representation of $P^{-\mu}_{\lambda}(x)$, valid for $\mathfrak{Re \mu}>\mathfrak{Re \lambda}>-1$ and $x>1$ is given by \cite{Bateman:100233}
\be\label{Pmulambda}
P^{-\mu}_{\lambda}(x) = \frac{(x^2 - 1)^{\mu/2}}{2^\lambda \Gamma(\mu - \lambda)\Gamma(\lambda +1)}\int\limits^{\infty}_0 \frac{(\sinh t)^{2\lambda +1}}{(x + \cosh t)^{\mu + \lambda +1}}dt \ .
\ee
Applying this to $P^{-\mu}_{\lambda}(\mu x)$ we get
\be
P^{-\mu}_{\lambda}(\mu x) = \frac{\mu^{-\lambda -1 }x^{-\lambda -1 } (1 -\frac{1}{\mu^2 x^2})^{\mu/2}}{2^\lambda \Gamma(\mu - \lambda)\Gamma(\lambda +1)}\int\limits^{\infty}_0 \frac{(\sinh t)^{2\lambda +1}}{(1 + \frac{\cosh t}{\mu x})^{\mu + \lambda +1}}dt \nonumber\ .
\ee
We can obtain the $\mu\rightarrow\infty$ limit with the help of the known asymptotics
\beqa\label{asympt}
&&(1 + y\epsilon)^{1/\epsilon} = e^y + \mathcal{O}(\epsilon)\ , \nonumber\\
&&\Gamma(z) = \sqrt{\frac{2\pi}{z}}\left(\frac{z}{e}\right)^z \left(1 + \mathcal{O}\left(\frac{1}{z}\right)\right) \ .
\eeqa
It is straightforward to get
\be\label{asympt1}
\Gamma(\mu +1)P^{-\mu}_{\lambda}(\mu x) = \frac{x^{-\lambda -1 }}{2^\lambda \Gamma(\lambda +1)}\int\limits^{\infty}_0 e^{-\frac{\cosh t}{x}}{(\sinh t)^{2\lambda +1}}dt + \mathcal{O}\left(\frac{1}{\mu}\right)\ ,
\ee
where we used $\frac{\Gamma(\mu +1)}{\Gamma(\mu - \lambda)} = \mu^{\lambda + 1}\left(1 + \mathcal{O}\left(\frac{1}{\mu}\right)\right)$,  which follows from (\ref{asympt}).

The asymptotic result (\ref{asympt1}) should be compared to the integral representation for $K_\nu (x)$ (valid for the same range of the parameter and argument) \cite{Bateman:100234}
\be
K_\nu (x) = \frac{\sqrt{\pi}x^{\nu }}{2^\nu \Gamma(\nu +\frac{1}{2})}\int\limits^{\infty}_0 e^{-x{\cosh t}}{(\sinh t)^{2\nu}}dt \nonumber\ .
\ee
It is immediately seen that by taking $\lambda = \nu - \frac{1}{2}$ we have
\be
\Gamma(\mu +1)P^{-\mu}_{\nu - \frac{1}{2}}(\mu x) = \sqrt{\frac{2}{\pi}}\sqrt{\frac{1}{x}} K_\nu \left(\frac{1}{x}\right)  + \mathcal{O}\left(\frac{1}{\mu}\right) \nonumber\ .
\ee
Upon identifying $x$ with $\frac{1}{2\beta z}$ and $\mu$ with $\frac{2}{\alpha}\kappa$, this result shows that the $R_{1-}$ solution in (\ref{solutions}) has the correct commutative limit (\ref{cmtvsln}). Incidentally, this also shows that to get the correct limit, $R_{1-}$ should be normalized by a factor of $\Gamma\left(\frac{2}{\alpha}\kappa +1\right)$. As the result, the full exact noncommutative solution in the free massive case takes the form
\be
\hat \Phi \;\sim \; \Gamma\left(\frac{2}{\alpha}\kappa +1\right) e^{i\beta\hat t}\,P_{\nu-\frac 12 }^{-\frac 2\alpha\kappa}\left(\frac{\hat r}{\alpha  \beta }\right)\,e^{i\beta\hat t}\label{zpttoth} \ .
\ee

Another consistency check of (\ref{zpttoth}) can be done by observing that near the boundary ($r\rightarrow \infty$, where now $r$ should be though as a symbol of $\hat r$) of noncommutative $AdS_2$, the space effectively becomes commutative \cite{Pinzul:2017wch,deAlmeida:2019awj}. Then the near-boundary behaviour of (\ref{zpttoth}) should reproduce the know result for the commutative field \cite{Freedman:1998tz}.

From \cite{gradshteyn2007} the asymptotic expansion for $P^{-\mu}_{\nu - 1/2}(x)$ as $x\rightarrow\infty$ is
\beqa
P^{-\mu}_{\nu - 1/2}(x)&=&\biggl\{\frac{2^{\nu - 1/2}\Gamma(\nu )}{\sqrt{\pi}\Gamma(\mu+\nu + 1/2)}x^{\nu - 1/2}\nonumber \\
&&+\frac{\Gamma(-\nu)}{2^{\nu + 1/2}\sqrt{\pi}\Gamma(\mu - \nu + 1/2)}x^{-\nu - 1/2}\biggr\}\Bigl(1+{\cal O}(x^{-2})\Bigr)\ ,\nonumber \\ &&(\nu - 1/2)\ne\pm \frac 12,\pm \frac 32,...\ .\nonumber\label{ohpttoth}
\eeqa
The two terms in the braces are the dominant ones when $\nu - 1/2$ and $-\nu - 1/2$ are greater than both $\nu - 5/2$ and $-\nu - 5/2$. Then for $-1<\nu<1$ and $\nu\ne 0$ (the Breitenlohner-Freedman (BF) bound further restricts to  $0<\nu<1$), defining $\Delta_\pm=\frac 12\pm\nu$, we obtain
\be
P_{-\Delta_- }^{-\frac 2\alpha\kappa}\left(\frac{r}{\alpha  \beta }\right)\rightarrow\frac{2^{-\Delta_- }\Gamma(\frac 12-\Delta_- )}{\sqrt{\pi}\Gamma(\Delta_+ +{\frac 2\alpha\kappa})}\left({\alpha  \beta }{ z}\right)^{\Delta_- }+\frac{2^{-\Delta_+}\Gamma(\Delta_- -\frac 12)}{\sqrt{\pi}\Gamma(\Delta_- +{\frac 2\alpha\kappa})}\left({\alpha  \beta }{ z}\right)^{\Delta_+ }\;,\nonumber \label{bbfLegfn}
\ee
as $z=\frac{1}{r}\rightarrow 0$ (here $z$ should be thought as the symbol of $\hat z$), which agrees with the asymptotic behavior of the commutative scalar field.\cite{Freedman:1998tz}

\subsection{Massless case}

The massless case corresponds to $m=0$, $\alpha\ne 0$, i.e. $\nu =\frac{1}{2}$. We can easily evaluate the integral (\ref{Pmulambda}) for $\lambda = \nu -\frac{1}{2} = 0$
\be\label{Pmu0}
P^{-\mu}_{0}(x) = \frac{(x^2 - 1)^{\mu/2}}{\Gamma(\mu )}\int\limits^{\infty}_0 \frac{\sinh t}{(x + \cosh t)^{\mu  +1}}dt = \frac{1}{\Gamma(\mu +1)}\left(\frac{x-1}{x+1}\right)^{\mu/2}\ .
\ee
(The branch-cut is taken in the usual way as to exclude the points $x=\pm 1$.) Actually, it is easily seen that in the massless case, two of the solutions in (\ref{solutions}) are allowed, $R_{1\pm}$, since passing from one to another is equivalent to replacing $\beta$ by $-\beta$. Using (\ref{Pmu0}), we see that the general massless solution is given by an arbitrary combination of (as before $\hat z = \hat{r}^{-1}$)
\be
e^{i\beta \hat t}\left(\frac {1\pm\alpha\beta\hat z}{1\mp\alpha\beta\hat z}\right)^{\frac\kappa\alpha}e^{i\beta\hat  t}\ .\label{mslssln}
\ee
Note that the factor of $\Gamma\left(\frac{2}{\alpha}\kappa +1\right)$ in (\ref{zpttoth}) exactly cancels out.

The result (\ref{mslssln}) should be compared with \cite{Pinzul:2021cjz} where exact solutions for the massless case were studied directly. There it was found that $\hat\Phi$ is an arbitrary combination (if one does not require reality) of operator functions $F_{\pm}$
\be
\hat\Phi = a F_{+}(\Xi_+) + b F_{-}(\Xi_-) \nonumber\ ,
\ee
where $\Xi_\pm := \pm\kappa\hat z + i \hat t$ and $a$ and $b$ are complex coefficients. We should show that this is consistent with (\ref{mslssln}).

Let us choose the upper sign in (\ref{mslssln}) (the other case is treated identically). Then using (\ref{hatthatzinv}) we get
\be
e^{i\beta \hat t}\left(\frac {1 + \alpha\beta\hat z}{1 -\alpha\beta\hat z}\right)^{\frac\kappa\alpha}e^{i\beta\hat  t} = \left({1 + 2\alpha\beta\hat z}\right)^{\frac\kappa\alpha}e^{2i\beta\hat  t}\nonumber\ .
\ee
We will show that
\be\label{massless}
\left({1 + 2\alpha\beta\hat z}\right)^{\frac\kappa\alpha}e^{2i\beta\hat  t} = e^{2\beta(\kappa\hat z + i\hat  t)}\equiv e^{2\beta\Xi_+}\ ,
\ee
which would prove the that the solutions are equivalent.

From (\ref{hatthatzinv}) follows the commutator $[\hat t, \hat z] = i\alpha\hat{z}^2$. Then the Baker–Campbell–Hausdorff formula (or rather the related Zassenhaus formula) gives
\be\label{BCH}
e^{2\beta(\kappa\hat z + i\hat  t)}e^{-2i\beta\hat  t} = f(\hat z)
\ee
for some function $f$ of only $\hat z$, which depends on $\beta$ as a parameter. Taking the derivative of (\ref{BCH}) with respect to $\beta$, we find
\be\label{fequation}
-2\alpha\hat{z}^2 f'(\hat z) + 2 \kappa\hat z f(\hat z) = \frac{\partial}{\partial\beta}f(\hat z) \ ,
\ee
where $f'(\hat z) = \frac{\partial}{\partial z}f( z)|_{z=\hat z}$ as usual. The solution should satisfy the condition $f(\hat z)|_{\beta = 0} = 1$. It is easy to see that
\be
f(\hat z) = \left({1 + 2\alpha\beta\hat z}\right)^{\frac\kappa\alpha} \nonumber
\ee
satisfies (\ref{fequation}) as well as the condition for $\beta = 0$. In this way (\ref{massless}) is established uniquely, and the proof that the exact solutions from \cite{Pinzul:2021cjz} are identical to the massless limit of (\ref{zpttoth}) is complete.

\section{Discussion}\label{Discussion}
The fact that $ncAdS$ leaves all the isometries of $AdS$ intact has always suggested that the free theory is exactly solvable. Yet, in the earlier works on $ncAdS$ \cite{Pinzul:2017wch,deAlmeida:2019awj} only perturbative solutions of free massless \cite{Pinzul:2017wch} and massive \cite{deAlmeida:2019awj} scalar field theories were constructed. In a recent article \cite{Pinzul:2021cjz} we obtained the general form of free massless scalar and fermionic theories. In the work presented here, using a completely different approach, we extended this result to the case of a massive scalar field. We believe that the extension to massive fermionic fields can also be done with the help of a map between the solutions in scalar and fermionic theories, which was found in \cite{Pinzul:2021cjz}.

The results obtained here are very important because knowledge of the exact (in the noncommutative parameter) free solutions is a necessary starting point for the perturbative treatment of interactions. In the presence of an interaction, there are two parameters - the noncommutativity parameter, $\alpha$, and the coupling constant, $\lambda$. Without the exact free solution, one has to resort to a perturbative analysis in \textit{both} parameters. This was the approach adopted in \cite{deAlmeida:2019awj}. Since only limited use of the isometry is made in this approach, the calculations are very involved and not intuitive. As a consequence, all the results were obtained only up to $\alpha^2$ order. Now having the exact noncommutative solution we hope to significantly improve the construction of the perturbation theory in a manner that will be exact in the noncommutative parameter. To make progress in this direction, one has to generalize the notion of Green function to the fully noncommutative setting. This problem is quite difficult both technically and conceptually and is the subject of our current research.

Another importance of the result of the present work is its relation to nonperturbative checks of the $AdS_2/CFT_1$ correspondence. More details on this is discussed in given in \cite{Pinzul:2022}.

\bibliographystyle{ws-rv-van}


\end{document}